\documentclass[conference]{IEEEtran}
\IEEEoverridecommandlockouts
\usepackage{cite}
\usepackage{amsmath,amssymb,amsfonts}
\usepackage{graphicx}
\usepackage{textcomp}
\usepackage{setspace}
\usepackage{nccmath}
\DeclareMathOperator*{\argmin}{argmin}

\usepackage{ntheorem}
\usepackage{tikz}
\usepackage{algorithm}
\usepackage[noend]{algorithmic}

\newcommand{\pars}[1]{\left(#1\right)}

\newcommand{\mat}[1]{\mathbf{#1}}
\newcommand{\R}{\mathbb{R}}

\newtheorem*{mle}{Maximum Likelihood Estimators}

\def\BibTeX{{\rm B\kern-.05em{\sc i\kern-.025em b}\kern-.08em
    T\kern-.1667em\lower.7ex\hbox{E}\kern-.125emX}}
\begin{document}

\title{
Semi-supervised Nonnegative Matrix Factorization for Document Classification*\\
\thanks{JH and DN were partially supported by NSF DMS $\#2011140$ and NSF BIGDATA $\#1740325$. JH is also partially supported by NSF DMS \#2111440. ES was supported by the Moore-Sloan Foundation. Funding from ICERM and the NSF-AWM ADVANCE grant initiated the collaboration.}
}

\author{\IEEEauthorblockN{Jamie Haddock}
\IEEEauthorblockA{\textit{Dept. of Mathematics} \\
\textit{Harvey Mudd College}
}
\and
\IEEEauthorblockN{Lara Kassab}
\IEEEauthorblockA{\textit{Dept. of Mathematics} \\
\textit{Colorado State Univ.}
}
\and
\IEEEauthorblockN{Sixian Li}
\IEEEauthorblockA{\textit{Dept. of Mathematics} \\
\textit{Univ. of Illinois Urbana-Champaign}
}
\and
\IEEEauthorblockN{Alona Kryshchenko}
\IEEEauthorblockA{\textit{Dept. of Mathematics} \\
\textit{Cal. State Univ. Channel Islands}
}
\and
\IEEEauthorblockN{Rachel Grotheer}
\IEEEauthorblockA{\textit{Dept. of Mathematics} \\
\textit{Wofford College}
}
\and
\IEEEauthorblockN{Elena Sizikova}
\IEEEauthorblockA{\textit{Center for Data Science} \\
\textit{New York Univ.}
}
\and
\IEEEauthorblockN{Chuntian Wang}
\IEEEauthorblockA{\textit{Dept. of Mathematics} \\
\textit{Univ. of Alabama}
}
\and
\IEEEauthorblockN{Thomas Merkh}
\IEEEauthorblockA{\textit{Dept. of Mathematics} \\
\textit{Univ. of Cal., Los Angeles}
}
\and
\IEEEauthorblockN{R.\ W.\ M.\ A.\ Madushani}
\IEEEauthorblockA{\textit{Dept. of Infectious Diseases} \\
\textit{Boston Medical Cent.}
}
\and
\IEEEauthorblockN{Miju Ahn}
\IEEEauthorblockA{\textit{Dept. of EMIS} \\
\textit{Southern Methodist Univ.}
}
\and
\IEEEauthorblockN{Deanna Needell}
\IEEEauthorblockA{\textit{Dept. of Mathematics} \\
\textit{Univ. of Cal., Los Angeles}
}
\and
\IEEEauthorblockN{Kathryn Leonard}
\IEEEauthorblockA{\textit{Dept. of Computer Science} \\
\textit{Occidental College}
}
}

\maketitle

\begin{abstract}
We propose new semi-supervised nonnegative matrix factorization (SSNMF) models for document classification and provide motivation for these models as maximum likelihood estimators. The proposed SSNMF models simultaneously provide both a topic model and a model for classification, thereby offering highly interpretable classification results. We derive training methods using multiplicative updates for each new model, and demonstrate the application of these models to single-label and multi-label document classification, although the models are flexible to other supervised learning tasks such as regression. We illustrate the promise of these models and training methods on document classification datasets (e.g., 20 Newsgroups, Reuters).
\end{abstract}


\section{Introduction}
Frequently, one is faced with the problem of performing a classification task on high-dimensional data which contains redundant information.  One such task is \emph{document classification} in which one assigns a set of categorical labels to documents based upon their contents~\cite{borko1963automatic, berry2009document}.  Document data is often represented using a \emph{bag-of-words} model, where the dimensionality of the representation of each document is linear in the number of unique words used in the document corpus and thus can be extremely large~\cite{harris1954distributional}. A common approach is to first apply a dimensionality-reduction technique (e.g., PCA \cite{pearson1901liii}), and then train a model for the classification task on the new, learned
representation of the data.  One problematic aspect of this two-step approach is that the learned representation of the data may provide ``good" fit, but could suppress data features which are integral to classification~\cite{guyon2003introduction}.
For this reason, supervision-aware dimensionality-reduction models have become increasingly important in data analysis. Such models aim to use supervision in the process of learning the lower-dimensional representation, or even learn this representation alongside the classification model~\cite{wang2014role,blei2003latent,pritchard2000inference}.

In this work, we propose new semi-supervised nonnegative matrix factorization (SSNMF) formulations which provide a dimensionality-reducing topic model and a model for a supervised learning task.
Our contributions are:
\begin{itemize}
    \item we motivate these proposed SSNMF models and that of~\cite{lee2009semi} as maximum likelihood estimators (MLE) given specific models of uncertainty in the observations;
    \item we derive multiplicative updates for the proposed models that allow for missing data and partial supervision; and
    \item we perform experiments on real data which illustrate the promise of these models in both topic modeling and supervised learning tasks
    relative to the performance of other relevant classifiers (e.g. Multinomial Naive Bayes).
\end{itemize}

\subsection{Notation}\label{sec:notation}
Our models make use of two matrix similarity measures.  The first is the standard Frobenius norm, $\|\mat{A} - \mat{B}\|_F$.  
The second is the \emph{information divergence} or I-divergence, a measure defined between nonnegative matrices $\mat A$ and $\mat B$,
\begin{equation}\label{def}
    D(\mat{A} \| \mat{B}) = \sum_{i,j} \bigg( \mat{A}_{ij} \log{\frac{\mat{A}_{ij}}{\mat{B}_{ij}}} - \mat{A}_{ij} + \mat{B}_{ij}\bigg),
\end{equation}
where $D(\mat{A} \| \mat{B}) \geq 0$ with equality if and only if $\mat{A} = \mat{B}$~\cite{lee2001algorithms}.
 
In the following, $\mat{A}/\mat{B}$ indicates element-wise division, $\mat A \odot \mat B$ indicates element-wise multiplication, and $\mat A \mat B$ denotes standard matrix multiplication.  We denote the set of non-zero indices of a matrix by $\text{supp}(\mat A) := \{(i,j) : \mat A_{ij}\ne 0\}$.  When an $n_1\times n_2$ matrix is to be restricted to have only nonnegative entries, we write $\mat A\geq 0$ and $\mat A\in \mathbb{R}^{n_1 \times n_2}_{\ge 0}$. We let $\mathbf{1_k}$ denote the length-$k$ vector consisting of ones, $ \mathbf{1_k} =  \begin{bmatrix}  1 , \cdots 1 \end{bmatrix}^\top \in \R^k$, and similarly $\mathbf{0_k}$ denotes the vector of all zeros, $\mathbf{0_k} =  \begin{bmatrix}  0 , \cdots 0 \end{bmatrix}^\top \in \R^k$.

We let $\mathcal{N}\pars{z \middle| \mu,\sigma^2}$ denote the Gaussian density function for a 
random variable $z$ with mean $\mu$ and variance $\sigma^2$, 
and $\mathcal{PO}\pars{z \middle| \nu}$ denotes the Poisson density function for 
a random variable $z$ with nonnegative intensity parameter $\nu$.
\subsection{Preliminaries}
\label{sec:preliminaries}
In this section, we give a brief overview of the NMF and SSNMF methods.

\subsubsection*{Nonnegative Matrix Factorization}
Given a nonnegative matrix $\mat X \in \mathbb{R}^{n_1 \times n_2}_{\ge 0}$ and a target dimension $r \in \mathbb N$, NMF decomposes $\mat X$ into a product of two low-dimensional nonnegative matrices.
The model seeks $\mat A$ and $\mat S$ so that $\mat X \approx \mat A \mat S$,
where $\mat A \in \mathbb{R}^{n_1 \times r}_{\ge 0}$ is called the dictionary matrix and $\mat S\in \mathbb{R}^{r \times n_2}_{\ge 0}$ is called the representation matrix.

Several formulations for this nonnegative approximation, $\mat X \approx \mat A \mat S$, have been studied~\cite{cichocki2009nonnegative,lee1999learning,lee2001algorithms,yang2011kullback}; 
e.g.,
\begin{equation}\label{eq:energy frobenius}
\argmin \limits_{\mat A\geq 0,\mat S\geq 0} \|\mat X - \mat A \mat S\|_F^2 \;\text{ and }
\argmin \limits_{\mat A\geq 0, \mat S\geq 0} D(\mat X\| \mat A \mat S),
\end{equation}
where $D(\cdot\|\cdot)$ is the information divergence defined in~\eqref{def}.
In what follows, we refer to the left formulation of \eqref{eq:energy frobenius} as $\|\cdot\|_F$-NMF and the right formulation of \eqref{eq:energy frobenius} as $D(\cdot\|\cdot)$-NMF. We refer the reader to~\cite{cichocki2009nonnegative} for discussions of similarity measures and generalized divergences (where information divergence is a particular case), and~\cite{li2012fast,sra2006generalized} for generalized nonnegative matrix approximations with Bregman divergences.

\subsubsection*{Semi-supervised NMF}
SSNMF is a modification of NMF to jointly incorporate a data matrix and a (partial) class label matrix. 
Given a data matrix $\mat X \in \R_{\ge 0}^{n_1 \times n_2}$ and a class label matrix $\mat Y \in \R_{\ge 0}^{k \times n_2}$, $(\|\cdot\|_F,\|\cdot\|_F)$-SSNMF is defined by 
\begin{equation} \label{eqn:ssnmf}
   \argmin\limits_{\mat A, \mat S, \mat B\geq 0} \underbrace{\|\mat W \odot(\mat X - \mat A \mat S)\|_F^2}_\text{Reconstruction Error} + \lambda \underbrace{\|\mat L \odot(\mat Y - \mat B \mat S)\|_F^2}_\text{Classification Error},
\end{equation}
where $\quad \mat A \in \R^{n_1 \times r}_{\ge 0},\, \mat B \in \R^{k \times r}_{\ge 0},\, \mat S \in \R^{r \times n_2}_{\ge 0}$, and the regularization parameter $\lambda >0$ governs the relative importance of the supervision term~\cite{lee2009semi}. 
The binary weight matrix $\mat W$ accommodates  missing data by indicating observed and unobserved data entries.
Similarly, $\mat L \in \R^{k \times n_2}$ is a weight matrix that indicates the presence or absence of a label.  
Multiplicative updates have been previously developed for SSNMF for the Frobenius norm, and the resulting performance of clustering and classification is improved by incorporating data labels into NMF~\cite{lee2009semi}.

\subsection{Related Work}\label{sec:related work}
In this section, we describe related work most relevant to our own.  This is not meant to be a comprehensive study of these areas.  

\subsubsection*{Statistical Motivation for NMF} 
The most common discrepancy measures for NMF $\|\cdot\|_F$-NMF and $D(\cdot\|\cdot)$-NMF correspond to the MLE given an assumed latent generative model and a Gaussian and Poisson model of uncertainty, respectively~\cite{cemgil2008bayesian,favaro20073,virtanen2008bayesian}. In \cite{cemgil2008bayesian,virtanen2008bayesian}, the authors go further towards a Bayesian approach, introduce application-appropriate prior distributions on the latent factors, and apply \emph{maximum a posteriori} (MAP) estimation.
Additionally, under certain conditions, it is shown that $D(\cdot\|\cdot)$-NMF is equivalent to probabilistic latent semantic indexing~\cite{ding2008equivalence}.

\subsubsection*{Dimension Reduction and Learning} 
There has been much work developing dimensionality-reduction models that are supervision-aware.
Semi-supervised clustering makes use of known label information or other supervision \emph{and} the data features while forming clusters~\cite{basu2002semi,klein2002instance,wagstaff2001constrained}. These techniques generally make use of label information in the cluster initialization or during cluster updating via must-link and cannot-link constraints; empirically, these approaches are seen to increase mutual information between computed clusters and user-assigned labels~\cite{basu2002semi}.
Semi-supervised feature extraction makes use of supervision information in the feature extraction process~\cite{fukumizu2004dimensionality,sheikhpour2017survey}.  These approaches are generally \emph{filter}- or \emph{wrapper}-based approaches, and distinguished by their underlying supervision type~\cite{sheikhpour2017survey}.

\subsubsection*{Semi-supervised and Joint NMF}
Since the seminal work of Lee et al.~\cite{lee2009semi}, semi-supervised NMF models have been studied in a variety of settings. The works~\cite{chen2008non,fei2008semi,jia2019semi} propose models which exploit cannot-link or must-link supervision.  In~\cite{cho2011nonnegative}, the authors introduce a model with information divergence penalties on the reconstruction and on supervision terms which influence the learned factorization to approximately reconstruct coefficients learned before factorization by a support-vector machine (SVM).
Several works~\cite{jia2004fisher,xue2006modified,zafeiriou2006exploiting} propose a supervised NMF model that incorporates Fisher discriminant constraints into NMF for classification.
Furthermore, joint factorization of two data matrices, like that of SSNMF, is described more generally and denoted Simultaneous NMF in~\cite{cichocki2009nonnegative}.

\subsection{Overview of Proposed Models}\label{sec:overview}
We propose two SSNMF formulations for document classification, both of which utilize information divergence on the first (data reconstruction) term.  This is a natural choice since many representations of document data (e.g., bag-of-words, n-grams, etc.) correspond to counts of word patterns in the data and are naturally modelled by Poisson distribution(s), which leads to the information divergence in the MLE model~\cite{cemgil2008bayesian,virtanen2008bayesian,hien2020algorithms}. Our proposed models accept document data $\mat X \in \mathbb{R}^{n_1 \times n_2}_{\ge 0}$, supervision matrix as $\mat Y \in \mathbb{R}^{k \times n_2}_{\ge 0}$, and target dimension $r$; we denote the models as $(D(\cdot,\cdot),\|\cdot\|_F)$-SSNMF,
\begin{equation} \label{eqn:divfrossnmf}
   \argmin\limits_{\mat A, \mat S, \mat B\geq 0} \underbrace{D(\mat W \odot \mat X, \mat W \odot \mat A \mat S)}_\text{Reconstruction Error} + \lambda \underbrace{\|\mat L \odot(\mat Y - \mat B \mat S)\|_F^2}_\text{Classification Error},
\end{equation}
and $(D(\cdot,\cdot),D(\cdot,\cdot))$-SSNMF,
\begin{equation} \label{eqn:divdivssnmf}
   \argmin\limits_{\mat A, \mat S, \mat B\geq 0} \underbrace{D(\mat W \odot \mat X, \mat W \odot \mat A \mat S)}_\text{Reconstruction Error} + \lambda \underbrace{D(\mat L \odot \mat Y, \mat L \odot \mat B \mat S)}_\text{Classification Error}.
\end{equation}

In each model, the matrix $\mat A \in \mathbb{R}_{\ge 0}^{n_1 \times r}$ provides a basis for the lower-dimensional space, $\mat S \in \mathbb{R}_{\ge 0}^{r \times n_2}$ provides the coefficients representing the projected data in this space, and $\mat B \in \mathbb{R}_{\ge 0}^{k \times r}$ provides the supervision model which predicts the targets given the representation of points in the lower-dimensional space.  We allow for missing data and labels or confidence-weighted errors via the data-weighting matrix $\mat W \in \mathbb{R}^{n_1 \times n_2}_{\ge 0}$ and the label-weighting matrix $\mat L \in \mathbb{R}^{k \times n_2}_{\ge 0}$. Each resulting joint-factorization model is defined by the error functions applied to the reconstruction and supervision factorization terms.

\section{SSNMF Models: Motivation and Methods}
\label{sec: I-SSNMF method}

In this section, we present a statistical MLE motivation of several variants of the SSNMF model, introduce the general semi-supervised models, and provide a multiplicative updates method for each variant.  %
While historically the focus of SSNMF studies have been on classification~\cite{lee2009semi}, 
we highlight that this joint factorization model can be applied quite naturally to 
regression tasks.

\subsection{Maximum Likelihood Estimation}\label{subsec:MLE}

In this section, we demonstrate that specific forms of our proposed variants of SSNMF are maximum likelihood estimators for given models of uncertainty or noise in the data matrices $\mat X$ and $\mat Y$.  These different uncertainty models have their likelihood function maximized by 
different error functions chosen for reconstruction and supervision errors, $R$ and $S$.  We summarize these results next; each MLE derived is a specific instance of a general model discussed in Section~\ref{subsec:propmodels}, in~\cite{lee2009semi}, or in~\cite{haddock2020semi}.  As mentioned previously, models which make use of the information-divergence objective are a natural choice since many representations of document data (e.g., bag-of-words, n-grams, etc.) are naturally modelled by Poisson distribution(s), which leads to the information divergence in the MLE model~\cite{cemgil2008bayesian,virtanen2008bayesian,hien2020algorithms}.

\begin{mle}
Suppose that the observed data $\mat X$ and supervision information $\mat Y$ have entries given as the sum of random variables, 
\begin{equation*}
    \mat X_{\gamma,\tau} = \sum_{i=1}^r x_{\gamma,i,\tau} \; \text{ and } \; \mat Y_{\eta,\tau} = \sum_{i=1}^r y_{\eta,i,\tau},
\end{equation*} 
and that the set of $\mat X_{\gamma,\tau}$ and $\mat Y_{\eta,\tau}$ are statistically independent conditional on $\mat A, \mat B$, and $\mat S$.
\begin{enumerate}
\itemsep-0.9em
\item When $x_{\gamma,i,\tau}$ and $y_{\eta,i,\tau}$ have distributions 
\begin{equation*}
    \mathcal{N}\pars{x_{\gamma,i,\tau} \middle| \mat A_{\gamma,i} \mat S_{i,\tau}, \sigma_1} \text{ and } \; \mathcal{N}\pars{y_{\eta,i,\tau} \middle| \mat B_{\eta,i} \mat S_{i,\tau}, \sigma_2}
\end{equation*} respectively, the maximum likelihood estimator is 
\begin{equation*}\label{fro-fro-MLE}
  \argmin_{\mat A, \mat B, \mat S \ge 0} \;\|\mat X - \mat A \mat S\|_F^2 + \frac{\sigma_1}{\sigma_2} \|\mat Y - \mat B \mat S\|_F^2.  
\end{equation*}


\item When $x_{\gamma,i,\tau}$ and $y_{\eta,i,\tau}$ have distributions $$\mathcal{N}\pars{x_{\gamma,i,\tau} \middle| \mat A_{\gamma,i} \mat S_{i,\tau}, \sigma_1} \text{ and } \mathcal{PO}\pars{y_{\eta,i,\tau} \middle| \mat B_{\eta,i} \mat S_{i,\tau}}$$ respectively, the maximum likelihood estimator is
$$\argmin_{\mat A, \mat B, \mat S \ge 0} \|\mat X - \mat A \mat S\|_F^2 +   2r\sigma_1 D(\mat Y\| \mat B \mat S).$$ \label{fro-div-MLE}

\item When $x_{\gamma,i,\tau}$ and $y_{\eta,i,\tau}$ have distributions $$\mathcal{PO}\pars{x_{\gamma,i,\tau} \middle| \mat A_{\gamma,i} \mat S_{i,\tau}} \text{ and } \mathcal{N}\pars{y_{\eta,i,\tau} \middle| \mat B_{\eta,i} \mat S_{i,\tau},\sigma_2}$$ respectively, the maximum likelihood estimator is 
$$\argmin_{\mat A, \mat B, \mat S \ge 0} D(\mat X\| \mat A \mat S) + \frac{1}{2r\sigma_2} \|\mat Y - \mat B \mat S\|_F^2.$$ \label{div-fro-MLE} 
\item When $x_{\gamma,i,\tau}$ and $y_{\eta,i,\tau}$ have distributions $$x_{\gamma,i,\tau} \sim \mathcal{PO}\pars{x_{\gamma,i,\tau} \middle| \mat A_{\gamma,i} \mat S_{i,\tau}} \text{ and } \mathcal{PO}\pars{y_{\eta,i,\tau} \middle| \mat B_{\eta,i} \mat S_{i,\tau}}$$ respectively, the maximum likelihood estimator is 
$$\argmin_{\mat A, \mat B, \mat S \ge 0} D(\mat X\| \mat A \mat S) + D(\mat Y\| \mat B \mat S).$$ \label{div-div-MLE}
\end{enumerate}
\end{mle}
\vspace{-0.5cm}

We note that \ref{div-div-MLE} follows from \cite{cemgil2008bayesian,favaro20073,virtanen2008bayesian}, but the others are distinct from previous MLE derivations due to the difference in the 
distributions assumed on data $\mat X$ and supervision $\mat Y$.  

\subsection{Multiplicative Updates}\label{subsec:propmodels}

The multiplicative updates method for all methods can be derived as follows \cite{lee2009semi}.
Suppose that the gradient of the objective function $F$ with respect to one of the variables $\Theta$ has a decomposition that is of the form: 
\[\nabla_\Theta F = [\nabla_\Theta F]^+ - [\nabla_\Theta F]^-, \]
where $[\nabla_\Theta F]^+ >0$ and $[\nabla_\Theta F]^->0$.
Then multiplicative update for $\Theta$ has the form
\[ \Theta \leftarrow \Theta \odot \frac{[\nabla_\Theta F]^-}{[\nabla_\Theta F]^+}.\]
We provide multiplicative updates for all three methods.  The pseudocodes for these methods are provided in~\cite{haddock2020semi}.

Implementation of these methods and code for experiments is available in the Python package \texttt{SSNMF}~\cite{SSNMFpackage}. 
Finally, we note that the behavior of these models and methods are dependent on the hyperparameters $r$, $\lambda$, and $N$.  One can select the parameters according to \emph{a priori} information or use a heuristic 
selection technique; 
we use both and indicate selected parameters and method of selection.

\subsection{Classification Framework} \label{sec:classification}
Here we describe a framework for using any of the SSNMF models for classification tasks. 
Given training data $\mat X_{\text{train}}$ (with any missing data indicated by matrix $\mat W_{\text{train}}$) and labels $\mat Y_{\text{train}}$, and testing data $\mat X_{\text{test}}$ (with unknown data indicated by matrix $\mat W_{\text{test}}$), 
we first train our $(R(\cdot \| \cdot),S(\cdot \| \cdot))$-SSNMF model 
to obtain learned dictionaries $\mat A_{\text{train}}$ and $\mat B_{\text{train}}$, where $R(\cdot \| \cdot)$ and $S(\cdot \| \cdot))$ denote specific metrics. We then use these learned matrices to obtain the representation of test data in the subspace spanned by $\mat A_{\text{train}}$, $\mat S_{\text{test}}$, and the predicted labels for the test data $\mat Y_{\text{test}}$.  

\noindent\textbf{Single-label Classification. } This process is: 
\begin{enumerate}
    \item Compute $\mat A_{\text{train}}, \mat B_{\text{train}}, \mat S_{\text{train}}$ as \[ \argmin\limits_{\mat A, \mat B, \mat S\geq 0} R(\mat W_{\text{train}} \odot \mat X_{\text{train}}, \mat W_{\text{train}} \odot \mat A\mat S) + \lambda S(\mat Y_{\text{train}}, \mat B\mat S).\]

    \item Solve $ \mat S_{\text{test}} = \argmin\limits_{\mat S\geq 0} R(\mat W_{\text{test}} \odot \mat X_{\text{test}}, \mat W_{\text{test}} \odot \mat A_{\text{train}}\mat S).$%
    
    \item Compute predicted labels as $\hat{\mat Y}_{\text{test}} = \text{label}(\mat B_{\text{train}}\mat S_{\text{test}})$, where $\text{label}(\cdot)$ assigns the largest entry of each column to 1 and all other entries to 0.
\end{enumerate}
In step 1, we compute $\mat A_{\text{train}}, \mat B_{\text{train}}$, and $\mat S_{\text{train}}$ using implementations of the multiplicative updates methods described above.  In step 2, we use either a nonnegative least-squares method (if $R = \|\cdot\|_F$) or one-sided multiplicative updates only updating $\mat S_{\text{test}}$ (if $R = D(\cdot \| \cdot)$). 
We note that this framework is significantly different than the classification framework proposed in~\cite{lee2009semi}; in particular, we use the classifier $\mat B$ learned by SSNMF, rather than independent SVM trained on the SSNMF-learned lower-dimensional 
representation to allow for an additional layer of interpretability.

\noindent\textbf{Multi-label Classification. } This framework generalizes to multi-label classification simply.  One first applies only steps (1) and (2) of the process above, forms $\hat{\mat Y}_{\text{test}} = \mat B_{\text{train}}\mat S_{\text{test}}$, and then applies a thresholding technique to decide the set of predicted labels for each data point; values above the threshold correspond to predicted labels, and those below to unpredicted labels.  There are many ways to do this; we instead vary the threshold uniformly between the minimum and maximum output values for each data point and report the highest encountered model metric.

\section{Experimental Data and Results}
\label{sec: numerical experiments}
In this section, we quantitatively evaluate the proposed methods on several document classification datasets to illustrate the promise of SSNMF models in both topic modeling and classification.

\subsection{20 Newsgroups Data Experiments}\label{subsec:20newsdata}
We first present our experiment on a subset of the 20 Newsgroups dataset~\cite{20news}, summarized in Table~\ref{table:20news_data}, where highlight the advantages of our SSNMF models and framework over benchmark methods.

We compute the term frequency–inverse document frequency representation for the documents, treat the groups as classes and assign them labels, and treat the subgroups as (un-labeled) latent topics in the data.  
We compare to the linear Support Vector Machine (SVM) and Multinomial Naive Bayes (NB) (see e.g.,~\cite{manning2008introduction}) classifiers, where the groups are treated as classes.
We also apply SVM as a classifier to the low-dimensional representation obtained from the NMF models, where (for both NMF and SSNMF models) we consider rank equal to 13 reflecting the number of subgroups in the dataset.
We consider all SSNMF models with the training process described in Section~\ref{sec:classification} with the maximum number of iterations (number of multiplicative updates) $N = 50$; our stopping criterion is the earlier of $N$ iterations or relative errorbelow tolerance $tol$.
We select the hyperparameters $tol$ and $\lambda$ for the models by searching over different values and selecting those with the highest average classification accuracy on the validation set.

\begin{table}[tb]
\centering
\footnotesize

\caption{Subset of the 20 Newsgroups dataset~\cite{20news} consisting of 5 groups and 13 subgroups partitioned roughly according to subjects.}\label{table:20news_data}
{ \renewcommand{\arraystretch}{1.15}
\begin{tabular}{ l  l }
            \hline
            Groups &\qquad Subgroups \\
            \hline
            Computers &\qquad graphics, mac.hardware, windows.x\\
            Sciences &\qquad crypt(ography), electronics, space\\
            Politics &\qquad guns, mideast \\
            Religion &\qquad atheism, christian(ity)\\
            Recreation &\qquad autos, baseball, hockey\\
\hline
\end{tabular}}
\end{table}

\begin{table}[htb]
\centering
\caption{Mean (and std. dev.) of test classification accuracy for each of the models on the subset of the 20 Newsgroups dataset described in Table~\ref{table:20news_data}.}\label{table:class_accuracy}
{ \renewcommand{\arraystretch}{1.15}
\footnotesize
\begin{tabular}{ |l|c| } 
\hline
Model & Class. accuracy \% (sd) \\
\hline
$(\|\cdot\|_F,\|\cdot\|_F)$ & 79.37 (0.47)  \\ 
$(\|\cdot\|_F,D(\cdot\|\cdot))$  &79.51 (0.38)  \\ 
$(D(\cdot\|\cdot),\|\cdot\|_F)$ & \textbf{81.88} (0.44)  \\ 
$(D(\cdot\|\cdot),D(\cdot\|\cdot))$ & 81.50 (0.47)  \\
$\|\cdot\|_F$-NMF + SVM & 70.99 (2.71)  \\
$D(\cdot \| \cdot)$-NMF + SVM &  74.75 (2.50) \\
\hline
SVM &  80.70 (0.27)  \\
Multinomial NB  &  \textbf{82.28} \\
\hline
\end{tabular}}
\end{table}

\begin{table*}[htb]
 \resizebox{\textwidth}{!}{
    \centering
\begin{tabular}{ |c|c|c|c|c|c|c|c|c|c|c|c|c|} 
\hline
Topic 1 & Topic 2 & Topic 3 & Topic 4 & Topic 5 & Topic 6 & Topic 7 & Topic 8 & Topic 9 & Topic 10 & Topic 11 & Topic 12 & Topic 13 \\
\hline
would & game & god & x & would & game & players & people & would & one & israel & like & god \\
space & team & would & thanks & armenian & one & team & israel & chip & us & guns & anyone & people \\
government & car & one & anyone & one & like & car & gun & key & get & people & available & church \\
use & games & jesus & graphics & people & car & last & right & algorithm & could & gun & key & one \\
key & engine & think & know & fbi & baseball & year & government & use & like & well & probably & christians \\
\hline
\end{tabular}
}
\caption{Top keywords representing each topic of the $(D(\cdot\|\cdot),\|\cdot\|_F)$-SSNMF model referred to in Figure~\ref{fig:SSNMF_Model5}. We (qualitatively) observe for example that topic 5, topic 8, and topic 11 capture the subjects of Middle East and guns (``israel", ``government", ``gun"). 
All three topics are associated with class Politics; see Figure~\ref{fig:SSNMF_Model5}.
We also observe that topic 1 and topic 9 relate to electronics/cryptography. Both are associated to class Sciences.
}
\label{table:model5_keywords}
\end{table*} 

We report in Table~\ref{table:class_accuracy} the average test classification accuracy for each of the models over 11 trials.
We define the test classification accuracy as $\sum_{i=1}^{n} \delta(\mat Y_i, \hat{\mat Y}_i)/n$, where $\delta(u,v)= 1$ for $u=v$, and 0 otherwise, and where $\mat Y_i$ and $\hat{\mat Y}_i$ are true and predicted labels, respectively.
We observe that the accuracy of $(D(\cdot\|\cdot),\|\cdot\|_F)$-SSNMF is comparable to
Multinomial NB which performs classification in the high-dimensional space. 
Note that the SSNMF models, which provide both dimensionality-reduction and classification in that lower-dimensional space, do not suffer great accuracy loss which suggests that the simultaneously learned low-dimensional representation serves the classification task well.
The SSNMF framework provides an intermediate layer that allows for additional interpretability by representing the data points in the low-dimensional topics space, where we learn about the shared and discriminative topics between classes. This serves the purpose of topic modeling (dimensionality reduction and clustering) and classification. 
Further, we observe that $(D(\cdot\|\cdot),\|\cdot\|_F)$-SSNMF performs significantly better than $D(\cdot \| \cdot)$-NMF + SVM in terms of accuracy emphasizing the importance of learning simultaneously a linear classifier and a low-dimensional representation.

\begin{figure}[tb]
    \centering
    \includegraphics[width=0.7\columnwidth]{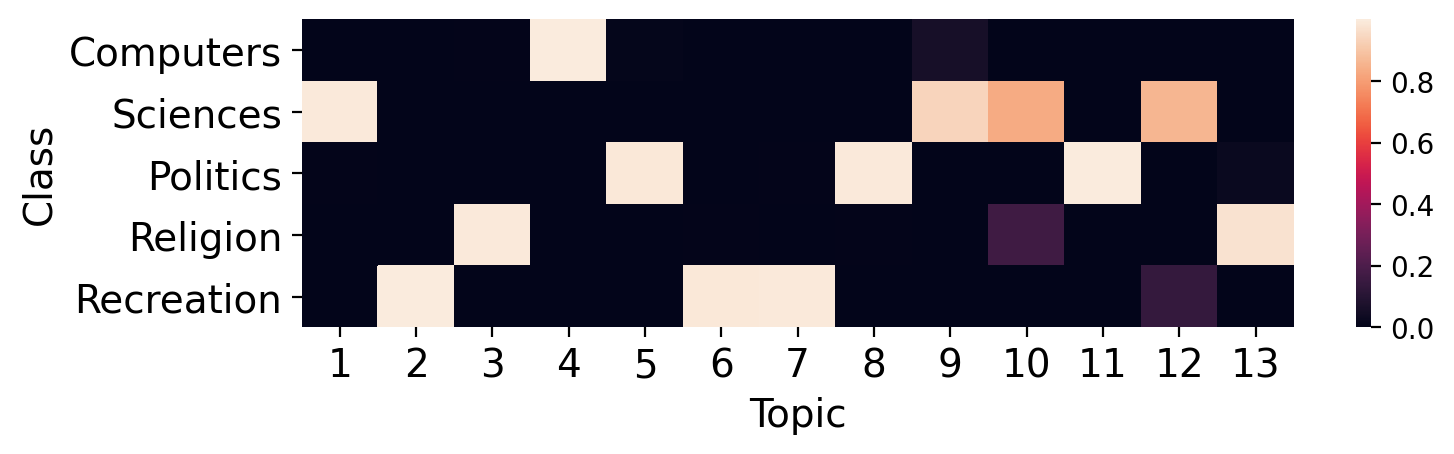}
    \caption{The normalized $\mat B_{\text{train}}$ matrix for the $(D(\cdot\|\cdot),\|\cdot\|_F)$ SSNMF decomposition corresponding to the median test classification accuracy equal to 81.78\% showcasing the topic distribution over classes.}
    \label{fig:SSNMF_Model5}
\end{figure}

Here, we consider the ``typical" decomposition for the $(D(\cdot\|\cdot),\|\cdot\|_F)$-SSNMF by selecting the decomposition corresponding to the median test classification accuracy.
We display in Figure~\ref{fig:SSNMF_Model5} the column-sum normalized $\mat B_{\text{train}}$ matrix of the decomposition, where each column illustrates the distribution of topic association to classes.
We display in Table~\ref{table:model5_keywords} the top 5 keywords (i.e. those that have the highest weight in topic column of $\mat A_{\text{train}}$) for each topic of the $(D(\cdot\|\cdot),\|\cdot\|_F)$-SSNMF of Figure~\ref{fig:SSNMF_Model5}.

\subsection{Reuters Data Experiments}\label{subsec:Reutersdata}

We next present our experiment on the Reuters Corpus~\cite{lewis2004rcv1}.  This corpus, which we download via \texttt{NLTK}, contains 10,788 news documents totaling 1.3 million words. The documents are classified into 90 classes (each document can have multiple labels and most do), and are grouped into two fixed sets, called ``training" and ``test."

We compute the term frequency–inverse document frequency representation for the documents, and apply the training process described in Section~\ref{sec:classification} (steps (1) and (2) for this multi-label classification task) with the maximum number of multiplicative updates iterations $N = 10$.
We set the hyperparameters $k = 200$ and $\lambda = 1$ for all models in this experiment.  In Table~\ref{table:micro-F1}, we present the mean and standard deviation of the micro-F1 score calculated on the test set over 100 trials. In each trial, we compute the matrix $\hat{\mat Y}_{\text{test}} = \mat B_{\text{train}}\mat S_{\text{test}}$ and vary the applied threshold for predicting labels uniformly (between data points) over the interval from the minimum entry to the maximum entry (in each data point).  That is, for each $\alpha \in [0,1]$, label $i$ is predicted for data point $j$ if the $(i,j)$th entry of $\hat{\mat Y}_{\text{test}} \ge \min \hat{\mat Y}_{\text{test}, j} + \alpha [\max \hat{\mat Y}_{\text{test}, j} - \min \hat{\mat Y}_{\text{test}, j}]$ where $\hat{\mat Y}_{\text{test}, j}$ is the $j$th column of $\hat{\mat Y}_{\text{test}}$.  In each trial, we compute this thresholded prediction for all $\alpha \in [0,1]$ and choose the largest micro-F1 score encountered.

\begin{table}[htb]
\centering
\caption{Mean (and std. dev.) of test micro-F1 score for each of the models on the subset of the Reuters dataset.}\label{table:micro-F1}
{ \renewcommand{\arraystretch}{1.15}
\footnotesize
\begin{tabular}{ |l|c| } 
\hline
Model & Micro-F1 Score \% (sd) \\
\hline
$(\|\cdot\|_F,\|\cdot\|_F)$ &   \textbf{41.59 (1.37)}\\ 
$(\|\cdot\|_F,D(\cdot\|\cdot))$  &   34.46 (1.76)\\ 
$(D(\cdot\|\cdot),\|\cdot\|_F)$ &   38.15 (2.08)\\ 
$(D(\cdot\|\cdot),D(\cdot\|\cdot))$ &   36.86 (2.51)\\
\hline
\end{tabular}}
\end{table}

\section{Conclusion} \label{sec:conclusion}
In this work, we have have proposed several SSNMF models, and have demonstrated that these models and that of~\cite{lee2009semi} are MLE in the case of specific distributions of uncertainty assumed on the data and labels.  We provided multiplicative update training methods for each model, and demonstrated the ability of these models to perform classification.

In future work, we plan to take a Bayesian approach to SSNMF by assuming data-appropriate priors and performing maximum \emph{a posteriori} estimation.  Furthermore, we will form a general framework of MLE models for exponential family distributions of uncertainty, and study the class of models where multiplicative update methods are feasible.

\section*{Acknowledgements}
The authors are appreciative of useful conversations with William Swartworth, Joshua Vendrow, and Liza Rebrova.

\bibliographystyle{plain}
\bibliography{main}

\end{document}